# NR-V2X Quality of Service Prediction Through Machine Learning with Nested Cross-Validation Scheme

İbrahim Yazici[1], and Emre Gures[2]
Türk Telekom R&D Department, Acıbadem, Istanbul, Turkey
Faculty of Electrical Engineering Czech Technical University in Prague, Prague, Czech Republic
[1]ibrahim.yazici@turktelekom.com.tr, [2]guresemr@cvut.cz

*Abstract*—The proliferation of connected vehicles and the advent of New Radio (NR) technologies have ushered in a new era of intelligent transportation systems. Ensuring reliable and low-latency communication between vehicles and their surrounding environment is of utmost importance for the success of these systems. This paper presents a novel approach to predict Quality of Service (QoS) in Vehicle-to-Everything (V2X) communications through nested cross-validation. Our methodology employs several machine learning (ML) methods to predict some QoS metrics, such as packet delivery ratio (PDR), and throughput, in NR-based V2X scenarios. In ML employment, nested cross-validation approach, unlike conventional cross-validation approach, prevents information leakage from parameter selection into hyperparameter selection, and this results in getting more robust results in terms of overfitting. The study utilizes real-world NR-V2X datasets to train and validate the proposed ML methods. Through extensive experiments, we demonstrate the efficacy of our approach in accurately predicting QoS parameters, even in dynamic and challenging vehicular environments. In summary, our research contributes to the advancement of NR-based V2X communication systems by introducing employment of ML methods with a novel approach for QoS prediction. The combination of accurate predictions through nested cross-validation not only enhances the reliability of communication in connected vehicles' landscape but also has a supportive role for stakeholders to make informed decisions for the optimization and management of vehicular networks.

## I. INTRODUCTION

At the forefront of transforming the transportation landscape, V2X technology (Vehicle-to-Everything) establishes a seamless connection between vehicles, enhancing road safety, efficiency, and sustainability. This groundbreaking paradigm transcends traditional communication models, enabling vehicles to exchange crucial information with infrastructure elements, vulnerable road users, and other vehicles within their vicinity. Pioneering protocols like IEEE 802.11p and Cellular Vehicle-to-Everything (C-V2X), standardized to fulfill diverse V2X application requirements, cater to varying quality of service (QoS) expectations, encompassing reliability, data rate, and latency [1], [2]. IEEE 802.11p, an evolutionary offshoot of the IEEE 802.11 standard, integrates Intelligent Transportation Systems (ITS) advancements and defines communication protocols for Dedicated Short-Range Communications (DSRC), enabling comprehensive vehicle-to-vehicle (V2V), vehicle-to-infrastructure (V2I), and vehicle-to-pedestrian (V2P) interactions [1].

The 3rd Generation Partnership Project (3GPP) further elevates the V2X paradigm with its C-V2X specification, introduced under Release 14 and leveraging cellular technology for V2X services [2]. C-V2X offers a range of benefits over IEEE 802.11p, including wider coverage, longer range, and higher throughput, making it particularly suitable for urban environments with dense traffic [3]. New Radio-V2X (NR-V2X), a 3GPP standard, emerges as a frontrunner for mission-critical safety applications, poised to empower advanced V2X use cases demanding exceptional communication speed, ultra-high reliability, and ultra-low latency, paving the way for groundbreaking concepts like remote driving, platooning, and extended sensor information exchange [3], [4].

Enabling predictive QoS is pivotal for enhancing V2X communication and enabling the automation of various applications. By accurately predicting network conditions, vehicular networks can proactively optimize performance, ensuring reliable data transmission. Additionally, predictive QoS empowers V2X applications to adapt their behavior, facilitating seamless and timely communication. A substantial body of research has investigated QoS prediction for V2X systems, with studies exploring various techniques and applications [5]–[9]. Ref. [5] investigates the use of supervised learning and ARIMA models to forecast the QoS of C-V2X communication in a typical urban setting, namely the Manhattan Grid. The authors gather a wide range of radio-related metrics, such as reference-signal-received quality/power (RSRQ/RSRP), the signal-to-noise ratio (SINR), the channel quality indicator (CQI), and past average throughput, as inputs to the models, ultimately assessing their predictive performance through accuracy and f1-score metrics. Ref. [6] examines the feasibility of employing machine learning (ML) techniques to predict uplink throughput for vehicular applications demanding high uplink requirements, such as tele-operated driving and local dynamic map update. The results demonstrate that deep neural network (DNN) and random forest (RF) consistently outperform logistic regression (LR) in terms of prediction accuracy. Ref. [7] proposes a (Long Short-Term Memory) LSTM-based QoS

This paper is funded by Türk Telekom, and authors of the paper thank to Türk Telekom for this financial support.

prediction framework for connected and automated vehicles. However, the complexity of the proposed approach may pose challenges for practical implementation in V2X applications. Ref. [8] presents a deep learning-based model for predicting CQI in V2I networks using received signal strength indication (RSSI) data. The model utilizes an LSTM network to capture the temporal dynamics of the channel and achieve high prediction accuracy compared to traditional ML models. Ref. [9] investigated the use of ML techniques to predict end-to-end (E2E) delay in vehicular communication systems. The authors employed a vehicle measurement campaign over three commercial long term evolution (LTE) networks to gather data on E2E delay, vehicle speed, SINR, RSRP, and RSSI. The results indicate that multilayer perceptron (MLP) and recurrent neural network (RNN) models outperformed other ML methods in classifying E2E delay.

As the literature on NR-V2X QoS prediction through ML is gleaned, to the best of our knowledge, it does not contain the ML deployment with nested cross-validation which prevents information leakage from parameter selection to model fitting, thereby generating more robust results against overfitting. ML employment with this novel approach does not only provide insight for researchers, it also presents a guideline in ML deployment to get more robust results in real-time V2X deployment which is an evolving field of wireless communication. Hence, the paper significantly makes contribution to the literature in this aspect.

The rest of the paper is organized as follows. Section II describes system model, and section III introduces proposed ML based prediction scheme. It is followed by section IV which provides performance evaluations for the proposed model. The paper is concluded in Section V.

## II. SYSTEM MODEL

To train our prediction models, we utilize a simulation-based approach, employing the publicly available dataset from Ref. [10]. This dataset was generated using Simulation of Urban MObility (SUMO), an open-source traffic simulator, to create realistic vehicle trajectories. Fifth-generation (5G)-Lena, an E2E open-source NR system-level simulator, was used to ensure seamless communication between the UE (user equipment) and the base station. The dataset encompasses four key parameters: modulation and coding scheme (MCS), distance-to-base station, SINR, and user datagram protocol (UDP) packet size. MCS was adjusted to optimize network performance, while distance-to-base station was calculated by utilizing coordinates extracted from both SUMO and the network simulator. SINR values were obtained upon simulation completion. To enhance data diversity, varying MCS and packet sizes were implemented during simulations and incorporated into the dataset without any pre-processing. This comprehensive dataset provided a valuable resource for training and evaluating prediction models for V2X communication under various scenarios.

## III. MACHINE LEARNING DEPLOYMENT

In this section we briefly give deployment details of ML models used in the paper. NR-V2X dataset used in the paper is introduced, and some data processing operations for the dataset are then given. Finally, used ML models are briefly introduced. In the ML deployments support vector regression (SVR), artificial neural network (ANN), gradient boosting regression (GBR), RF, light gradient boosting method (LGBM), categorical boosting regression (CBR) are utilized with nested cross-validation and their results are compared in terms of different metrics. In the comparisons of the ML model results, mean absolute error (MAE), root mean square error (RMSE) are utilized. In addition, R-squared (R2 score) metric for all the models, which shows how much variance in the inputs are explained by the output, is used for the comparisons of the models. A dataset, briefly explained in the previous section, is utilized, which contains several data recordings including SINR, packet size, distance-to-base station, throughput, MCS generated through SUMO traffic simulator and 5G-LENA network simulator, for ML deployment.

### A. Data Set and Data Pre-processing

In this paper, we utilize ML models to predict throughput and path loss values by using simulation generated distance-to-base station, packet data size, MCS and SINR values as inputs. The dataset is acquired through SUMO and 5G-Lena simulators from different scenarios in four different simulation spots, three of which is from İstanbul, and the other one is from Berlin. The dataset comprises of diverse data generated as a result of different scenario adjustments, such as using different base station placement, packet sizes, Tx power, vehicle location and Tx power, thereby creating a valuable and abundant data set for NR-V2X QoS prediction. The dataset is examined whether duplication is available, and it was spotted that the set contains missing values. After the examinations, duplicate values are removed, and missing values are removed from the set. As a result, the dataset used for ML deployment in this paper contains 8716 data samples. For ML methods, only ANN and SVR methods require normalization of features, so that we employ zero-mean normalization only for these ML methods in this study to enable faster convergence and getting stable results. At the end, we use a novel training-validation scheme, nested cross-validation, for ML methods. Unlike traditional cross-validation scheme, nested cross-validation distinctly perform parameter selection and model selection, thereby preventing information leakage between parameter selection and model selection. This ability offers preventing overfitting of dataset and producing more robust and reliable results with respect to the traditional scheme.

### B. Machine Learning Methods

In this paper, we use CBR, SVR, RF, GBR, ANN, and LGBM methods to predict throughput and PDR values. Even some advanced ML methods, such as LSTM, GRU etc. are used for V2X- and C-V2X-related problems, our dataset is more suitable for conventional ML method use due to its tabular data type rather than being kind of data suitable for deep learning methods. We thus employ conventional ML methods by using nested cross validation, thereby providing a new perspective for both research and practical applications. In nested cross-validation employment settings, we use 8-fold split in outer loop while using 6-fold split in inner loop.

*CBR:* Categorical Boosting Regression is one of the popular boosting methods in recent times proposed by Yandex researchers [11]. We use it with a hyperparameter setting which depth is in the range [4,5,7,10], minimum child samples are in the range [1, 4, 8, 16], and learning rates are in the range [0.01,0.03,0.09,0.1,0.5,0.9], and iterations are in the range [150,200].

*SVR:* Support Vector method is a kernel-based method [12] which separates a dataset into different classes by maximizing the margins between different class examples. Although it was originally proposed for classification task, it is used for regression tasks as well. It is called SVR when it is used for continuous values to find a hyperplane that best represent the data points in a continuous space. We use it with a hyperparameter setting which kernel type is set to radial basis function, C values are 1 and 3, epsilon values are 0.1 and 0.3.

*RF:* Random Forest is one of the prevalently used ensemble learning method that uses bagging strategy by employing many decision trees [13]. We use it with a hyperparameter setting which number of estimators are in the range of [50,100,200], minimum samples leaf numbers are in the range of [2,3,4,6].

*GBR:* Gradient Boosting method is also one of the prevalently used ML method for tabular type data, and it is based on constructing strong learners from weak learners, thereby eliminating the weakness of the weak learners [14]. We use it with a hyperparameter setting which number of estimators are in the range of [1,10,50,100,300,500,700], maximum depth is in the range of [1,3,5,7], and learning rate is in the range of [0.01,0.03,0.09,0.3].

*ANN:* Artificial Neural Network is another powerful ML method for both classification and regression tasks [15]. It processes the given input(s) utilizing activation functions between the layers of the network by projecting the processed data in different spaces in each layer, and performs its training with backpropagation method which results in adjusting weights and bias parameters of the network. We use ANN with a hyperparameter setting which hidden layer sizes are in the range [5,10,25], number of maximum iteration is in the range [500,1500,2500]. Rectified Linear Unit (ReLU) activation function is used in the hidden layers in the network.

*LGBM:* Light Gradient Boosting method [16] is another type of boosting method used in this paper. It similarly uses constructing strong learners from weak learners as GBR, however, it grows leaf-wise contrary to depth-wise growth of GBR. We use LGBM with a hyperparameter setting which learning rate is in the range [0.003, 0.006,0.009,0.01,0.03,0.06,0.09,0.3,0.6], number of estimators are in the range of [20,40,80,100], number of leaves are in the range of [10,15,20,25], colsample bytree is in the range of [0.7, 0.8, 0.9], and maximum bin is in the range of [75,150, 255, 510].

We split the data set into 8 folds in the outer loop, and into 6 folds in the inner loop of the nested cross validation. Accordingly, we utilize MAE and RMSE metrics for performance comparison of the predictions of throughput and PDR by ML methods used in the study. Moreover, R2 score is also used for the performance comparison of the ML methods.

TABLE I
THROUGHPUT AND PDR PREDICTION RESULTS

| Model | MAE | | RMSE | | R2 | |
|---|---|---|---|---|---|---|
| | Thr. | PDR | Thr. | PDR | Thr. | PDR |
| CBR | 0.210 | 4.551 | 0.625 | 12.771 | 0.951 | 0.921 |
| SVR | 1.089 | 28.7 | 1.741 | 43.602 | 0.627 | 0.076 |
| RF | 0.183 | 3.873 | 0.656 | 12.832 | 0.947 | 0.92 |
| GBR | 0.206 | 3.939 | 0.667 | 12.71 | 0.945 | 0.92 |
| ANN | 1.173 | 16.432 | 1.569 | 24.56 | 0.698 | 0.706 |
| LGBM | 0.238 | 4.753 | 0.661 | 13.065 | 0.946 | 0.917 |

## IV. RESULTS, PERFORMANCE COMPARSIONS AND DISCUSSIONS

In this section, we present the results of throughput and PDR predictions by the ML methods. The results in terms of MAE, RMSE and R2 score are given in Table I.

Some visuals related to the results in Table I are also illustrated by Figures 1-4.

As per the given results in Table I, in terms of MAE metrics, RF method achieves the best result in predicting throughput by 0.183. It is followed by GBR and CBR methods which their results are very close to each other as seen in Figure 1. On the other hand, RF achieves the best result in predicting PDR by 3.873 as well. It is again followed by GBR and CBR in order. LGBM strictly follows GBR and CBR both in the predictions of throughput and PDR as seen in Figure 2. ANN follows LGBM with a major performance difference for both of the predictions, and SVR yields the worst results in the predictions.

In terms of RMSE metrics, CBR method achieves the best result in predicting throughput by 0.625, and it is strictly followed by RF and GBR methods in order as seen in Figure 3. GBR method achieves the best result in predicting PDR by 12.71, it is strictly followed by CBR and RF methods in PDR prediction in order. LGBM again follows the predictions in the fourth rank with a moderate difference margin in the predictions. Similar to MAE results, ANN follows LGBM in both predictions as seen in Figure 4, and SVR method again produces the worst results among the methods.

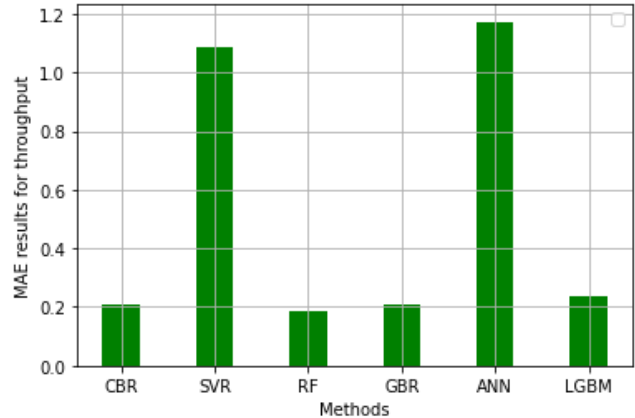

Fig. 1. MAE results for throughput predictions

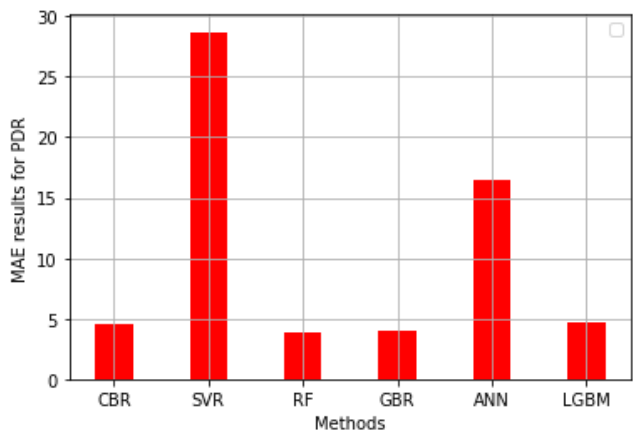

Fig. 2. MAE results for PDR predictions

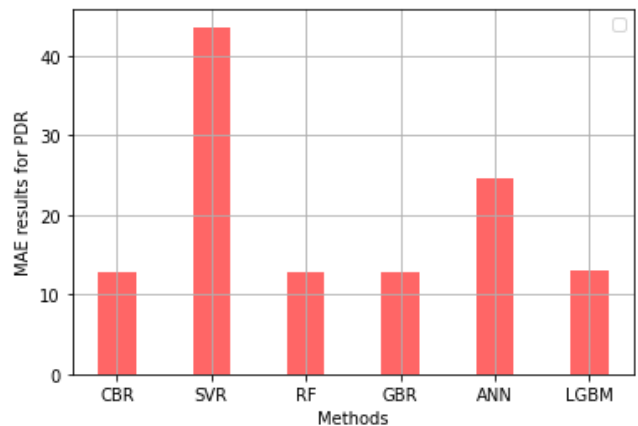

Fig. 4. RMSE results for PDR predictions

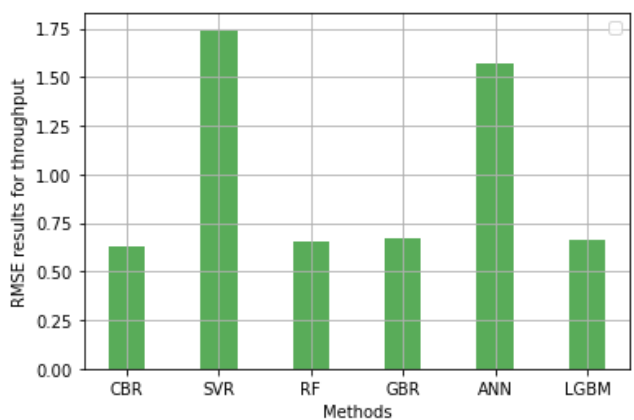

Fig. 3. RMSE results for T predictions

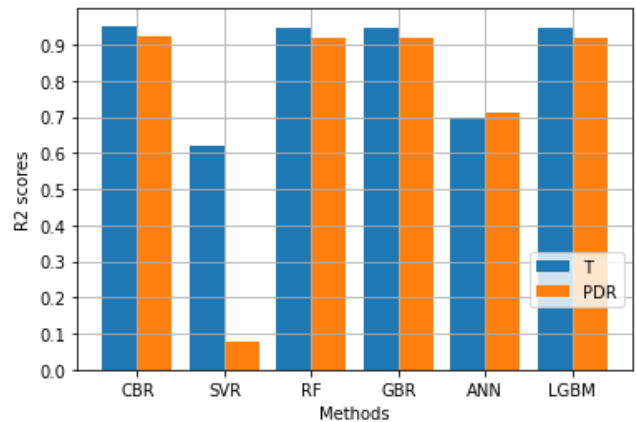

Fig. 5. R2 scores for the predictions

R2 score is a measure that indicates how much the variance of a dependent variable is represented by independent variable(s). It is also monitored for all the methods for the predictions, and R2 scores for the methods are illustrated in Figure 5. RF, GBR, CBR and LGBM R2 scores nearly yield the same results by around 0.95 for throughput model fitting. ANN performs worse than these four methods for throughput model fitting, and SVR performs the worst in this respect as seen from Figure 5. On the other hand, RF, GBR, CBR and LGBM R2 scores again yield nearly the same results in PDR model fitting by around 0.92, and they are followed by ANN. SVR apparently performs the worst when compared to the rest of the methods used in the paper.

When method-level analysis is conducted, it is seen RF method achieves superior results with respect to GBR and CBR in terms of MAE metrics. It yields better results than GBR and CBR, in order, in throughput prediction by 0.13 and 0.15 performance gains. In terms of RMSE metrics, CBR method yields better results than RF and GBR, in order, in PDR prediction by 0.05 and 0.07 performance gains. On the other hand, even GBR method yield the best result with respect to RF and CBR methods in PDR prediction, its performance gain is very minuscule. Overall, among the methods used in the paper, ensemble learning methods achieve the prevalent results with respect to ANN and SVR methods which has landmark importance in the context of ML methods in the literature. One reasoning about this fact may be that boosting and tree methods are more adept in tabular type of dataset with respect to the other ML methods. On the other hand, dataset size is one of the limitations of this study, and ANN might yield better results than the current one with a large dataset. Moreover, experiments with different parameter settings might produce for SVR to obtain challenging results with boosting and tree methods. In future research studies, more abundant dataset use to get challenging results leveraging different type of ML methods is considered as a future topic within NR-V2X scope.

## V. CONCLUDING REMARKS

In conclusion, this research focused on predicting NR-V2X QoS leveraging ML methods, specifically targeting throughput and PDR as key outputs. Leveraging inputs such as MCS, packet delay, SINR, and distance to the base station, a comparative analysis was conducted across various ML algorithms, including RF, GBR, LGBM, ANN and SVR.

A distinctive aspect of this study was the implementation of a novel cross-validation scheme known as nested cross-validation. This approach effectively mitigates information leakage from model selection into hyperparameter selection, enhancing the robustness of the ML models in comparison to traditional cross-validation methods. The evaluation of predictions was carried out through metrics such as MAE, RMSE, and R2 score. Notably, RF emerged as the top-performing method in throughput and PDR predictions, showcasing its effectiveness based on the MAE metric. GBR and CBR closely followed it, contributing to the overall efficiency of ensemble learning methods in this context. Furthermore, the study revealed that CBR excelled in throughput prediction according to the RMSE metric, while GBR outperformed in PDR prediction. Overall, ensemble learning methods consistently demonstrated favorable R2 scores compared to ANN and SVR.

While the findings underscore the overall efficiency of ML with the used approach, particularly ensemble learning, in predicting NR-V2X QoS, it is essential to acknowledge the limitation of the dataset size. Future research endeavors should prioritize expanding the dataset to provide a more comprehensive understanding of this pivotal area within 5G and beyond 5G (B5G). The deployment of ML with nested cross-validation not only advances the research landscape but also offers practical guidance for practitioners seeking robust and stable results for real-time applications in the evolving field of wireless communication.